\documentclass[prl,twocolumn,showpacs,preprintnumbers,amsmath,amssymb]{revtex4}

\usepackage{graphicx}
\usepackage{dcolumn}
\usepackage{bm}

\begin{document}


\title{Vortex Redistribution below the First-Order Transition Temperature 
in the $\beta$-Pyrochlore Superconductor KOs$_2$O$_6$} 

\author{T. Shibauchi,$^{1,2}$ M. Konczykowski,$^2$ C.~J. van der Beek,$^2$ 
R. Okazaki,$^1$ Y. Matsuda,$^{1,3}$ \\
J. Yamaura,$^3$ Y. Nagao,$^3$ and Z. Hiroi$^3$}

\affiliation{$^1$Department of Physics, Kyoto University,
Sakyo-ku, Kyoto 606-8502, Japan\\
$^2$Laboratoire des Solides Irradi\'es, Ecole Polytechnique, 91128 
Palaiseau cedex, France\\
$^3$Institute for Solid State Physics, University of Tokyo,
Kashiwa, Chiba 277-8581, Japan}

\date{\today}

\begin{abstract}
A miniature Hall sensor array was used to detect magnetic induction locally 
in the vortex states of the $\beta$-pyrochlore superconductor KOs$_2$O$_6$. 
Below the first-order transition at $T_{\rm p}\sim 8$~K, 
which is associated with a change in the rattling motion of K ions, 
the lower critical field and the remanent magnetization both show 
a distinct decrease, suggesting that the electron-phonon coupling 
is weakened below the transition. 
At high magnetic fields, the local induction shows an unexpectedly 
large jump at $T_{\rm p}$ whose sign changes with position 
inside the sample. 
Our results demonstrate a novel redistribution of vortices 
whose energy is reduced abruptly below the first-order transition 
at $T_{\rm p}$.

\end{abstract}

\pacs{74.25.Op, 74.25.Qt, 74.25.Ha, 74.25.Kc}

\maketitle

Owing to the competition between vortex interactions, 
thermal and quantum fluctuations, and the effect of quenched disorder, 
the physics of type-II superconductors involves a rich discussion
on the nature of ``vortex matter'' \cite{Blatter}. 
In recent years, much attention has been focused 
on vortex-lattice melting, which 
has been found to be a first-order phase transition 
in high-temperature cuprate superconductors 
\cite{Safar,Pastoriza,Zeldov,Schilling}. 
The first-order transition, vortex pinning, and 
surface barrier effects are at the origin of nontrivial spatial 
distributions of the vortex density inside superconductors, 
which have been the subject of intense research 
\cite{Soibel,Marchevsky,Brandt}.

Very recently, an intriguing first-order transition, completely 
distinct from vortex lattice melting, has been observed in the vortex state 
of the $\beta$-pyrochlore superconductor KOs$_2$O$_6$, that has a relatively
 high transition temperature $T_{\rm c} =$ 9.6~K \cite{YonezawaK,Hiroi}. 
In this system, a remarkable feature is the 
anharmonic ``rattling'' motion of K ions within 
an oversized Os-O atomic cage \cite{Kunes,Yamaura}; this 
is believed to be responsible for the unusual convex temperature dependence 
of the resistivity $\rho(T)$ in the normal state \cite{Hiroi}. 
The first-order transition is manifest through
a nearly field-independent peak in the specific heat 
at $T_{\rm p}\sim 8$~K \cite{Hiroi,Batlogg,Hiroi_p}. 
High-field transport measurements have revealed that 
the concave $\rho(T)$ at high temperatures changes to 
a $T^2$ dependence below the first-order 
transition at $T_{\rm p}$ \cite{Kasahara,Hiroi_p}, 
suggesting that the phonon spectrum responsible for the 
anomalous transport properties has a substantial change at $T_{\rm p}$. 
Recent measurements of thermodynamical quantities \cite{Hiroi_p}, 
NMR \cite{Yoshida}, photoemission \cite{Shimojima}, 
and penetration depth \cite{Bonalde,Shimono} suggest 
superconductivity with strong electron-phonon coupling in this system. 
The effect of the transition on vortex matter in KOs$_2$O$_6$ is of 
fundamental interest, as this gives one a sensitive probe of the 
change of the free energy at $T_{\rm p}$.

Here, by utilizing a micro-Hall probe magnetometry, we show that 
the vortex distribution changes in a drastic way below the 
first-order transition at $T_{\rm p}$. 
We observe a huge local induction jump, exceeding the global magnetization 
jump expected from thermodynamics by more than an order of magnitude. 
This results from a novel vortex redistribution driven by the transition. 
The formation of this new vortex distribution is 
a consequence of an abrupt decrease of the vortex line energy 
below $T_{\rm p}$, which 
is evident from the observed 
distinct changes of the lower critical field $H_{\rm c1}$. 

KOs$_2$O$_6$ single crystals were grown 
by the technique described elsewhere \cite{Hiroi}. 
Since it has been known that partial hydration diminishes the anomaly 
in specific heat at $T_{\rm p}$, special care was taken to keep the 
crystals in a dry atmosphere before the measurements. 
We use a sensitive Hall-sensor array tailored in a GaAs/AlGaAs 
two dimensional electron gas system. Each sensor has an active area of 
6$\times 6~\mu$m$^2$, and the center-to-center distance of neighboring 
sensors is 20~$\mu$m. A KOs$_2$O$_6$ crystal with dimensions  
$110 \times 270 \times 90~\mu$m$^3$ is placed on top of the array; 
the magnetic field $H$ is applied along the 90~$\mu$m direction 
by using 
a low-inductance 1.8-T superconducting magnet with a 
negligibly small remanent field. 

\begin{figure}
\includegraphics[width=90mm]{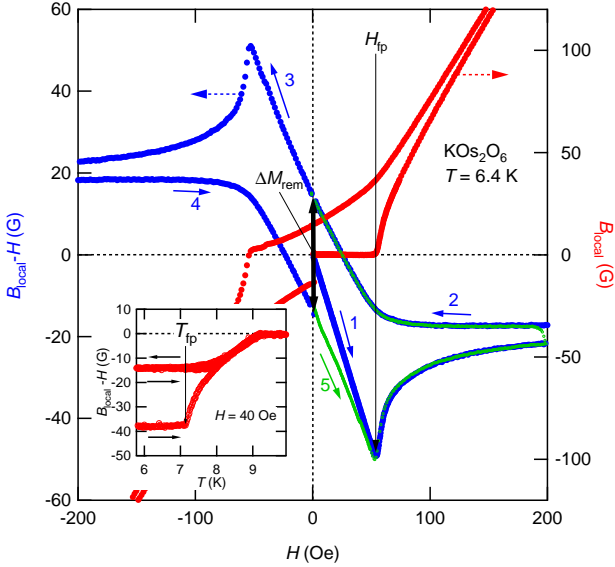}%
\caption{(color online). 
Local magnetic induction $B_{\rm local}$ (right axis) 
and magnetization (left axis) measured at 6.4 K near the sample center 
as a function of field $H$. 
The numbers and thin solid arrows indicate the order and direction 
of the field sweeps. 
At the first penetration field $H_{\rm fp}$, a sharp dip is observed. 
The remnant magnetization $\Delta M_{\rm rem}$ is defined by 
the hysteresis width at $H=0$~Oe (thick solid arrow). 
Inset: the temperature dependence of local magnetization ($H=40$~Oe) shows 
a clear kink at the first penetration temperature $T_{\rm fp}$. }
\end{figure}

{\it Lower Critical Field.}---
We first report on results of low-field magnetometry. 
In Fig.~1 we show the field dependence of local induction $B_{\rm local}(H)$ 
measured by the sensor placed close to the center of the crystal in the 
zero-field cooling (ZFC) condition at $T=6.4$~K. 
At very low fields, the sample is in the 
Meissner state $B_{\rm local}=0$~G. Above a first-penetration field 
$H_{\rm fp}=53$~Oe, vortices enter and the induction increases rapidly. 
By plotting the ``local magnetization'' defined by $B_{\rm local}-H$, 
the first penetration field is more clearly resolved, as a sharp dip 
with a diverging ${\rm d}B_{\rm local}/{\rm d}H$ at $H_{\rm fp}$. 
The dip (or peak) positions in the second loop coincide with those in 
the loop of first magnetization. This behavior strongly suggests 
that the small overall hysteresis is governed by geometrical (surface) 
barriers 
\cite{ZeldovPRL,Brandt} and the contribution of bulk pinning 
in this system is quite small. 
Then, the lower critical field $H_{\rm c1}$ can be 
determined from the expression for the first-penetration field 
of a superconducting bar, 
accounting for the demagnetization effect: 
\begin{equation}
H_{\rm fp}/H_{\rm c1} = \tanh\bigl(\sqrt{0.36\ b/a} \bigr), 
\label{eq:demag}
\end{equation}
where $b/a$ is aspect ratio of the bar, with the (perpendicular) 
field along the thickness $2b$ \cite{Brandt}. 

\begin{figure}
\includegraphics[width=90mm]{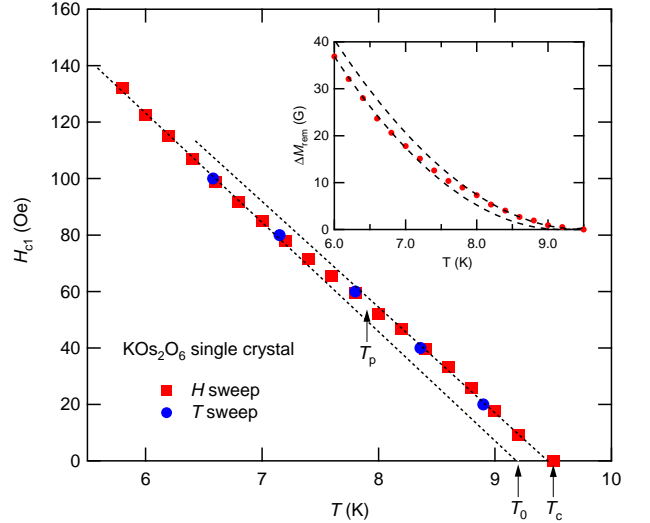}%
\caption{(color online). Lower critical field as a function of temperature 
in KOs$_2$O$_6$ obtained form the Hall sensor magnetometry. 
Inset: the temperature dependence of the remanent magnetization. 
The dotted and dashed lines represent linear and quadratic temperature 
dependences, respectively. 
}
\end{figure}

From Eq.~(\ref{eq:demag}), we evaluate $H_{\rm fp}/H_{\rm c1}=0.50$ 
and thus we obtain $H_{\rm c1}(T)$ as depicted in Fig.~2. 
We also observe a sharp kink at $T_{\rm fp}$ in the temperature 
dependence of the local magnetization in the ZFC condition as shown in 
the inset of Fig.~1. These $H$ and $T$--sweep measurements consistently 
yield a single $H_{\rm c1}(T)$ curve. Now it is clear that 
$H_{\rm c1}(T)$ below $T_{\rm p}\sim 8$~K is considerably lower than 
that extrapolated from higher temperature data, although the slope 
is not very different between above and below $T_{\rm p}$. 
Since the low temperature data extrapolates to a temperature 
$T_0=9.2$~K, clearly 
lower than $T_{\rm c}$, the relative decrease in $H_{\rm c1}$ immediately 
indicates that the effective transition temperature ({\it i.e.} the 
superconducting gap) is reduced below $T_{\rm p}$. 
This concurs with recent microwave penetration depth 
$\lambda(T)$-measurements \cite{Shimono}, 
in which a similar shift of the superfluid density $\lambda^{-2}(T)$ 
to a lower value is observed below $T_{\rm p}$. 
We also note that although a global magnetization measurement 
\cite{Schuck} gave a somewhat smaller value of $H_{\rm c1}$(5~K), 
the observed slope ${\rm d}H_{\rm c1}/{\rm d}T=-38$~Oe/K 
near $T_{\rm c}$ gives a very good agreement 
with the reported slope values of upper critical field 
${\rm d}H_{\rm c2}/{\rm d}T=-33$~kOe/K 
\cite{Shibauchi,Schuck} and the thermodynamic critical field 
${\rm d}H_{\rm c}/{\rm d}T=-570$~Oe/K 
through the thermodynamic relation
$H_{\rm c1}H_{\rm c2}=H_{\rm c}^2\ln \kappa$, 
where $\kappa\approx 80$ \cite{Hiroi_p,Shimono} 
is the Ginzburg-Landau parameter. 

A similar feature is seen in the temperature dependence of 
the remanent local magnetization $\Delta M_{\rm rem}$, 
defined by the hysteresis width at $H=0$~Oe [see Fig.~1], 
which also shows a clear change below $T_{\rm p}$. 
As shown in inset of Fig.~2, $\Delta M_{\rm rem}$ roughly follows 
quadratic temperature dependence, but 
the low temperature data again extrapolate to 
below $T_{\rm c}$. These results all indicate 
that below $T_{\rm p}$, the superconducting condensation energy,
the effective $T_{\rm c}$, and the superconducting gap become 
smaller than the value extrapolated downward from $T_{\rm c}$.
According to a recent theory \cite{Dahm}, 
an anharmonic damped phonon mode 
is responsible for the unusual $\rho(T)$, 
and its abrupt change from the convex temperature 
dependence above $T_{\rm p}$ to a $T^2$ dependence below 
indicates that the effective phonon frequency is increased dramatically 
below $T_{\rm p}$. 
In this view, the observed reduction of effective $T_{\rm c}$ 
implies that the electron-phonon coupling 
strength is weakened below the first-order transition. 

{\it Local Induction Jump.}---
Next, we turn to the higher field measurements up to 18~kOe. 
According to the Clausius-Clapeyron relation 
\begin{equation}
\Delta B = -4\pi 
\left( {{\rm d}T_{\rm p} \over {\rm d} H} \right) 
\left[ \Delta S  - \Delta V 
\left( \frac{{\rm d} T_{\rm p}} {{\rm d}p}\right)^{-1} \right], 
\label{eq:CC}
\end{equation}
the first order transition at $T_{\rm p}$ should be accompanied by 
a discontinuity $\Delta B$ of the global induction 
(here $\Delta S$ is the entropy change per unit volume 
at the transition, $\Delta V$ is the volume change of 
the sample, and $p$ is pressure). However, given the near-independence 
of $T_{\rm p}$ on field and no detectable volume change at $T_{\rm p}$ 
by the structural analysis \cite{XRD}, 
this jump should be modest. From the data of 
Ref.~\onlinecite{Hiroi_p}, $\Delta S = 360$ mJ/mol K 
and $dT_{\rm p}/dH \approx 0.7$~mK/kOe, the global induction 
jump at the transition should be $\sim 0.4$~G. 
Indeed, recent SQUID measurements give a close value 
$\Delta B\sim 0.5$~G 
at 2~T \cite{Hiroi_p}, 
which maintains the thermodynamic consistency. 

Surprisingly, however, the local induction $B_{\rm local}(T)$ 
reveals a much larger jump at $T_{\rm p}$. 
In Fig.~3(a) we plot the  
change of the local induction $\delta B_{\rm local}(T)$ 
relative to the normal state near the center of the sample. 
With decreasing temperature, $\delta B_{\rm local}$ becomes negative below 
$T_{\rm c}(H)$ and decreases down to $T_{\rm p}$. At $T_{\rm p}$, 
there is a jump with field-dependent magnitude, 
to the extent that $\delta B_{\rm local}(T)$ becomes positive at high fields. 
The magnitude $\Delta B_{\rm local}$ of the jump 
increases linearly with $H$ as shown in the inset; 
at $H=18$~kOe it reaches $\sim 8$~G, 
which is more than an order of magnitude larger 
than the global jump. 

\begin{figure}
\includegraphics[width=92mm]{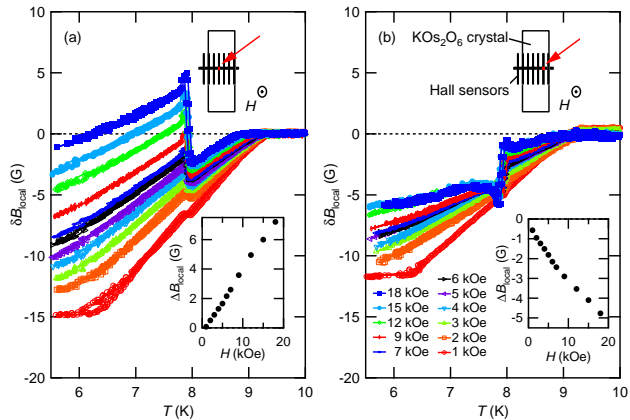}%
\caption{(color online). Temperature dependence of the local induction 
change $\delta B_{\rm local}$ with respect to the value in the normal 
state, for different applied fields near the crystal center (a) and 
away from the center (b). 
The insets show the magnitude of the jump  $\Delta B_{\rm local}$ 
at $T_{\rm p}$ as a function of $H$. 
}
\end{figure}

Most remarkably, the behavior of the $\Delta B_{\rm local}$--jump 
at $T_{\rm p}$ strongly depends on the position at which it is measured 
on the samples. Away from the center, the sign of the induction jump 
at high fields is reversed as shown in Fig.~3(b): with decreasing temperature 
$\delta B_{\rm local}(T)$ decreases abruptly below $T_{\rm p}$. 
Such a position dependence of the magnetization jump 
is not usually expected for a first-order phase transition. 
In the first-order vortex lattice melting transition, for example, 
the induction jump has the same sign everywhere in the sample \cite{Soibel}. 
This highlights the novelty of the present observation in KOs$_2$O$_6$. 
We also note that the observed behavior of $\delta B_{\rm local}(T)$ is 
almost reversible with temperature cycles, indicating that 
bulk pinning by defects does not play a significant role here.

\begin{figure}
\includegraphics[width=87mm]{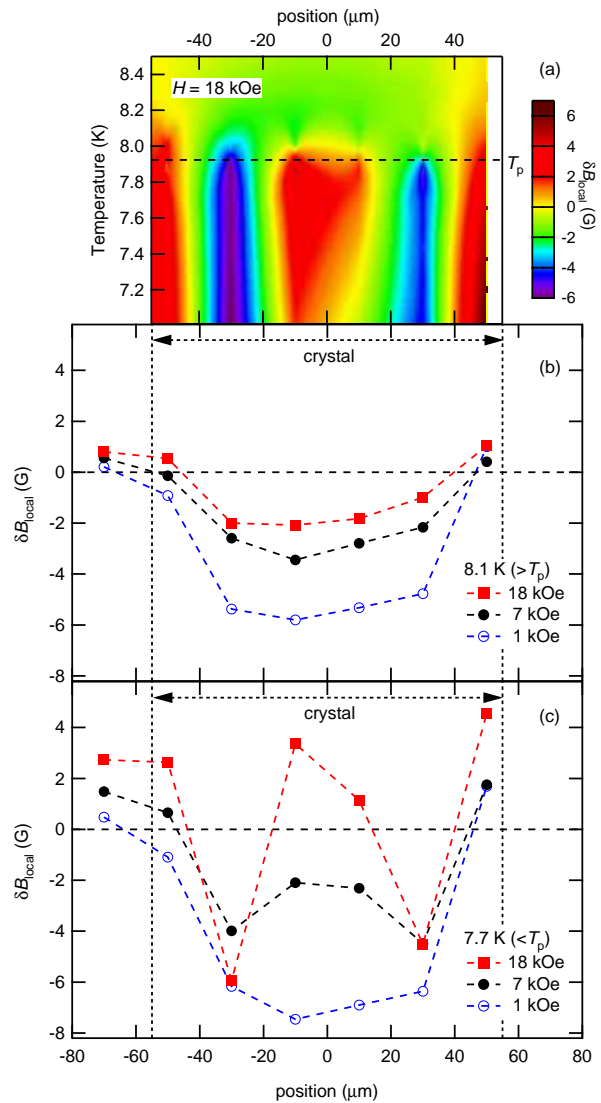}%
\caption{(color online). 
Local induction change $\delta B_{\rm local}$ 
as a function of position. (a) Contour image of $\delta B_{\rm local}$ 
inside the sample at 18 kOe, constituted from interpolations 
and plotted in the temperature-position plane. 
This highlights an abrupt change of the vortex distribution near $T_{\rm p}$. 
The position dependence of $\delta B_{\rm local}$ at several fields 
is plotted just above (b) and below $T_{\rm p}$ (c). 
The dotted lines indicate the approximate positions of the sample 
edges, and dashed lines are guides to the eye.}
\end{figure}

{\it Vortex Redistribution.}---
To see the anomalous jump at the transition more clearly, we 
plot the position dependence of $\delta B_{\rm local}$ in Fig.~4. 
Above $T_{\rm p}$ [Fig.~4(b)], the induction profile shows 
the standard behavior: 
$\delta B_{\rm local}$ is negative inside the sample and the difference 
between $B_{\rm local}$ and $H$ shrinks with increasing field. 
We note that the positive $\delta B_{\rm local}$ near the edges 
is due to 
the stray fields by the shielding supercurrent, which are 
expected for the sample with a finite demagnetizing factor. 
It should be also noted that the local induction $B_{\rm local}$ 
inside the sample except for the thin regions 
near the edges should be equal to $n_{\rm v}\Phi_0$, 
where $n_{\rm v}$ is the density of vortices and $\Phi_0$ 
is the flux quantum \cite{Zeldov}. 
Just below $T_{\rm p}$, the field distribution 
drastically changed especially at high fields [Fig.~4(c)]. 
Near the center, the induction 
jumps to a higher value, while away from the center, 
it jumps to a lower value. 
Correspondingly, the vortex density profile now 
has a characteristic dome-like shape 
with troughs away from the center. 
The overall shape remains unchanged 
down to low temperature [Fig.~4(a)]. A somewhat similar dome shaped field 
profile has been studied for thin high-$T_{\rm c}$ cuprate crystals 
at high temperature and low magnetic fields, where geometrical barriers
against flux entry govern the vortex distribution \cite{ZeldovPRL}. 
In contrast, our results are obtained in a field range much higher than 
$H_{\rm c}$, where such barrier effects may not be important. 

Next we discuss the possible mechanism of this novel vortex 
redistribution below $T_{\rm p}$. Since we experimentally 
observe the reduced $H_{\rm c1}$ 
below $T_{\rm p}$ [Fig.~2], the free energy of 
a single vortex per unit length 
\begin{equation}
\varepsilon = \Bigl({\Phi_0 \over 4\pi\lambda}\Bigr)^2\ln\kappa
= {\Phi_0 \over 4\pi}H_{\rm c1} 
\end{equation}
should also be smaller in the low-temperature phase. At the first-order 
transition, the low-$T$ phase and high-$T$ phase coexist, and the region 
of the low-$T$ phase is invested by ``cheaper'' vortices 
with smaller $\varepsilon$. Then vortices near the boundaries 
between the two phases should be attracted 
to the low-$T$ phase region. This mechanism promotes inhomogeneous vortex 
density: denser in the low-$T$ phase and sparse in the high-$T$ phase regions. 
The difference $\Delta n_{\rm v}$ in vortex density 
between the two phases is determined by the balance of 
energy gain $ \Delta n_{\rm v} \Delta \varepsilon $ and 
the kinetic energy loss $\sim ( \Delta n_{\rm v} )^{2} $ due to  
the net current tracing the phase boundary. 
As temperature is lowered, the whole sample is transformed to the 
low-temperature phase. However, the mutual repulsion between the shielding 
supercurrent near the crystal edges and this current 
``string'' surrounding the low-temperature phase nucleus prevents the 
dome-like flux distribution from relaxing.  In this way, we may have the 
dome-shaped vortex distribution resembling the situation discussed in the 
presence of the geometrical surface barriers \cite{ZeldovPRL}. 
Although a more quantitative theoretical investigation is 
necessary for the full understanding of the novel vortex redistribution, 
the energy difference per unit area between low and high-$T$ phases is 
given by $\Delta \varepsilon B/\Phi_0$, which may be responsible 
for the observed $H$-linear dependence of the jump height 
$\Delta B_{\rm local}$ [insets of Fig.~3]. 

In summary, by using the micro-Hall probe array, we have measured the local 
magnetization at the surface of the KOs$_2$O$_6$ 
single crystal presenting a first-order transition within the 
superconducting state. A number of anomalies associated with the transition 
have been observed. Below the first-order transition temperature $T_{\rm p}$, 
(i) the lower critical field is shifted down, (ii) the remanent magnetization 
shows a relative decrease as well, (iii) the local induction reveals 
huge jumps which depend on the position inside the sample in an 
unusual way, and (iv) there is an abrupt vortex redistribution into a 
flux dome, which we believe decorates the nucleating low-temperature 
phase. Our results indicate that the change in the rattling motion reduces 
the superconducting critical fields as well as the vortex energy.

We acknowledge fruitful discussion with T. Dahm, R. Ikeda, A.~I. Buzdin, 
and S. Fujimoto. 
This work was partly supported by Japan-France Integrated 
Action Program SAKURA from JSPS, 
and by Grants-in-Aid for Scientific Research of MEXT, Japan. 
T. S. is also grateful to the hospitality of the LSI people during 
his stay at Ecole Polytechnique.

\end{document}